\documentclass[superscriptaddress, preprintnumbers,amsmath,amssymb, aps, prd, floatfix, twocolumn]{revtex4}
\usepackage{mathrsfs}
\usepackage{graphicx}
\usepackage{dcolumn}
\usepackage{bm}
\renewcommand{\labelenumi}{(\roman{enumi})}
\usepackage{comment}
\usepackage{braket}
\usepackage{amsthm}
\usepackage{tikz}
\usepackage{color,soul}
\usetikzlibrary{positioning}
\usepackage{hyperref}
\hypersetup{
    colorlinks=false,
    linkcolor=blue,
    filecolor=magenta,      
    urlcolor=cyan,
}

\begin{document}

\preprint{CHIBA-EP-247, 2021.06.23}

\title{Reconstructing confined particles with complex singularities
}

\author{Yui Hayashi}
\email{yhayashi@chiba-u.jp}
\affiliation{
Department of Physics, Graduate School of Science and Engineering, Chiba University, Chiba 263-8522, Japan
}

\author{Kei-Ichi Kondo}
\email{kondok@faculty.chiba-u.jp}
\affiliation{
Department of Physics, Graduate School of Science and Engineering, Chiba University, Chiba 263-8522, Japan
}
\affiliation{
Department of Physics, Graduate School of Science, Chiba University, Chiba 263-8522, Japan
}

\begin{abstract}
Complex singularities have been suggested in propagators of confined particles, e.g., the Landau-gauge gluon propagator.
We rigorously reconstruct Minkowski propagators from Euclidean propagators with complex singularities.
As a result, the analytically continued Wightman function is holomorphic in the tube, and the Lorentz symmetry and locality are kept intact, whereas the reconstructed Wightman function violates the temperedness and the positivity condition.
Moreover, we argue that complex singularities correspond to confined zero-norm states in an indefinite metric state space.
\end{abstract}

\maketitle

\section{INTRODUCTION}

Color confinement, the absence of colored degrees of freedom from the physical spectrum, is an essential element of strong interactions.
Understanding this fact in the framework of relativistic quantum field theory (QFT) is a fundamental issue of particle and nuclear physics.

To investigate such fundamental aspects of strong interactions, the gluon, ghost, and quark propagators in the Landau gauge have been extensively studied by both lattice and continuum methods \cite{Alkofer:2000wg, Huber:2018ned, Maas13}.
Based on this progress, there has recently been an increasing interest in the analytic structures of the gluon, ghost, and quark propagators \cite{Alkofer:2003jj, SFK12, HFP14, Siringo16a, Siringo16b, DOS17, Lowdon17, Lowdon18, Lowdon:2018mbn, CPRW18, HK2018, DORS19, KWHMS19, BT2019, LLOS20, HK2020, Fischer-Huber, Falcao:2020vyr}.
In particular, \textit{complex singularities}, which are unusual singularities invalidating the K\"all\'en-Lehmann spectral representation \cite{spectral_repr_UKKL}, attract much attention.
In old literature \cite{Gribov78, Zwanziger89, Stingl85, HKRSW90, Stingl96, Dudal:2008sp}, e.g., for models motivated by the Gribov ambiguity, it was predicted that the gluon propagator in the Landau gauge has a pair of complex poles, which is a typical example of such singularities.

The recent studies done without assuming the K\"all\'en-Lehmann spectral representation, e.g., reconstructing the propagators from Euclidean data \cite{BT2019, Falcao:2020vyr}, modeling the propagators by the gluon mass \cite{TW10,TW11,PTW14,Siringo16a, Siringo16b, HK2018, HK2020}, and the ray technique of the Dyson-Schwinger equation \cite{SFK12,Fischer-Huber} consistently indicate the existence of complex singularities of the Landau-gauge gluon propagator.
Note that complex singularities were also observed using the ray technique in other models \cite{Maris:1991cb, Maris:1994ux, Maris:1995ns}.


Since complex singularities should never appear in propagators of observable particles, we can expect that they are connected to confinement. 
Thus, theoretical aspects of complex singularities are of crucial importance since they could provide some hints for a better understanding of a confinement mechanism.
Theoretical consequences of complex singularities have so far been discussed only heuristically.
For example, some argue that the appearance of complex singularities might imply nonlocality, e.g., \cite{Stingl85, HKRSW90, Stingl96}. Nevertheless, this argument is not fully convincing due to the use of the naive inverse Wick-rotation.
To our knowledge, a solid study on this subject is still lacking.

Therefore, we will scrutinize the reconstruction procedure, namely the reconstruction of Wightman functions, or the vacuum expectation values of the products of field operators, from Schwinger functions, or the Euclidean correlators, by the analytic continuation.

\begin{figure}[t]
  \begin{center}
  \begin{center}
   \includegraphics[width=0.9 \linewidth]{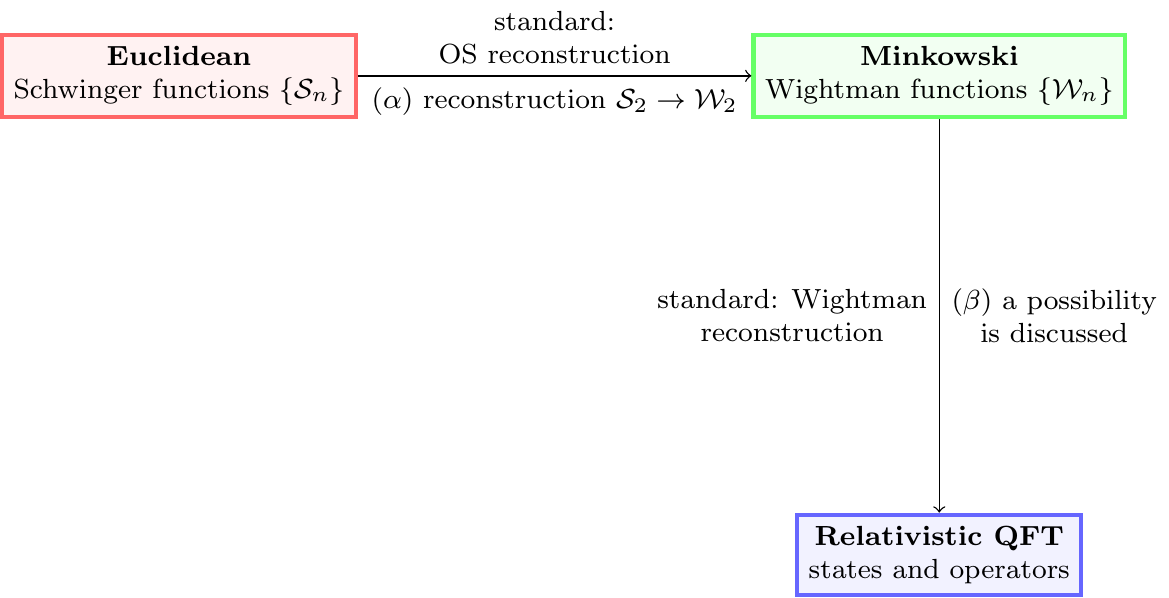}
  \end{center}

  \end{center}
   \caption{The standard reconstruction procedure and contents of our study consisting of ($\alpha$) and ($\beta$)}
    \label{fig:introduction}
\end{figure}

We consider formal aspects of complex singularities.
Figure \ref{fig:introduction} frames our study in the reconstruction procedure.
In the standard reconstruction procedure, one begins with a family of Schwinger functions satisfying Osterwalder-Schrader (OS) axioms \cite{OS73, OS75} and reconstructs a QFT based on the OS theorem and the Wightman reconstruction theorem (see Theorem 3-7 in \cite{Streater:1989vi}).
In our study, we start from two-point Schwinger functions in the presence of complex singularities violating some of the OS axioms as seen below.
We first reconstruct the Wightman function based on the holomorphy in ``the tube'' \cite{Streater:1989vi} according to the flow shown in Fig.~\ref{fig:introduction}.
We then examine general properties of the two-point Wightman function and discuss possible state-space structures.


In this paper, we illustrate a typical example of a propagator with complex singularities and give a sketch of main results, omitting mathematical subtleties.
We provide full details of their rigorous proofs and derivations in a longer version \cite{HK2021}.

\section{Setup and Main results}

We use the following notations:
The space of test functions with compact supports is denoted by $\mathscr{D}(\mathbb{R}^4)$, and that of rapidly decreasing test functions by $\mathscr{S}(\mathbb{R}^4)$.
Elements of these dual spaces $\mathscr{D}'(\mathbb{R}^4)$ and $\mathscr{S}'(\mathbb{R}^4)$ 
are called distributions and tempered distributions, respectively.
We use $x,y,\xi,\eta$ as elements of $\mathbb{R}^4$.
We write $\xi = (\vec{\xi}, \xi_4)$ for Euclidean space and $\xi = (\xi^0, \vec{\xi})$ for Minkowski or complexified space.
In accordance with \cite{Streater:1989vi}, $\mathbb{R}^4 - i V_+$ is called the tube, where $V_+$ denotes the (open) forward light cone 
\begin{align}
    V_+ := \{ (\eta^0,\vec{\eta}) \in \mathbb{R}^4 ~;~ \eta^0 > |\vec{\eta}|  \}.
\end{align}

\begin{figure}[t]
  \begin{center}
   \includegraphics[width=0.85 \linewidth]{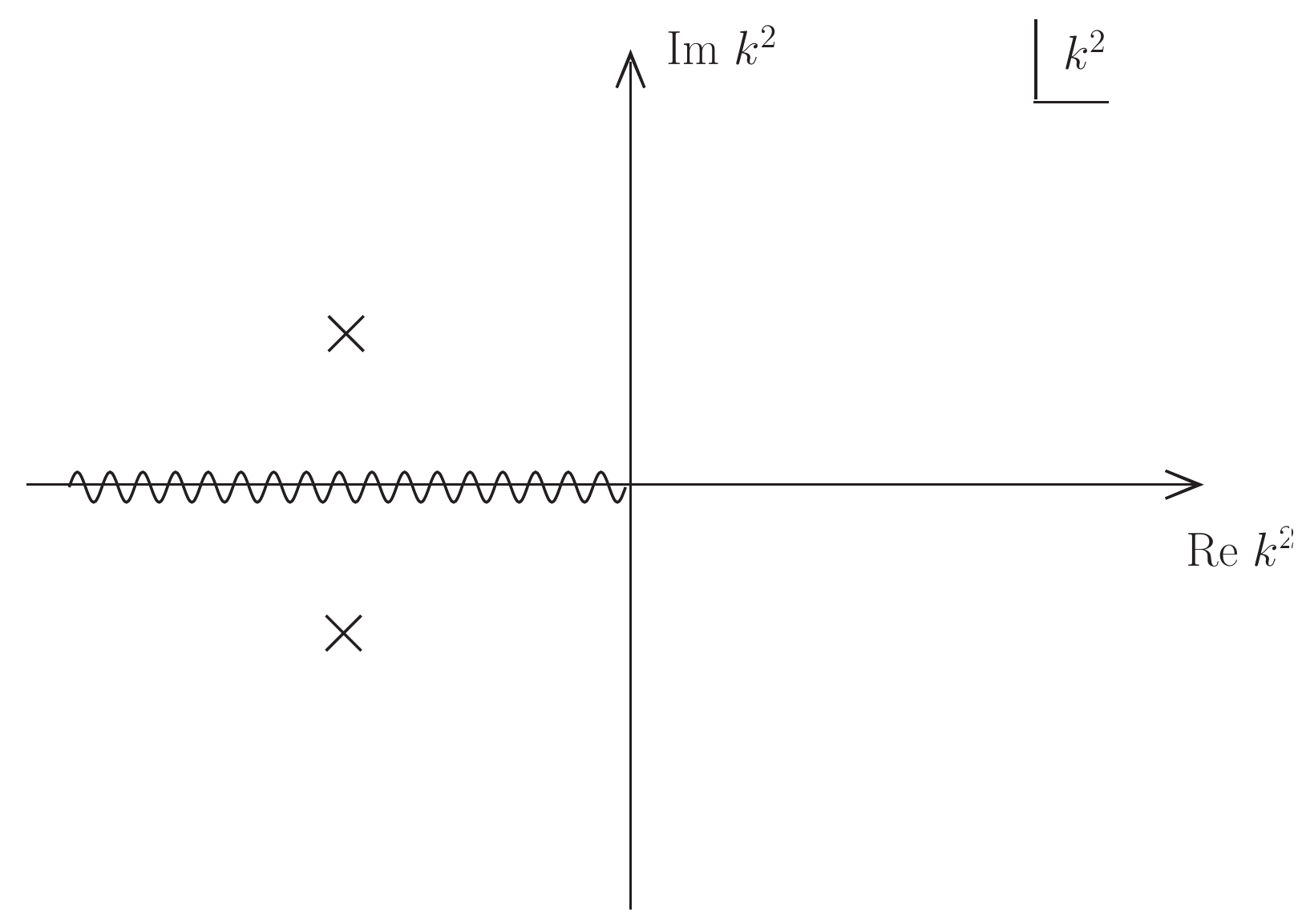}
  \end{center}
   \caption{
   Complex squared momentum $k^2$ plane of the propagator $D(k^2)$ with a pair of complex conjugate poles given in (\ref{eq:propagator_complex_poles}).
   }
    \label{fig:complex.pdf}
\end{figure}

For simplicity, we consider a two-point function for a scalar field $\phi(x)$.
In all the evidences for complex singularities mentioned above, the complex singularities appear in an analytically-continued Euclidean propagator on the complex squared momentum $k^2$ plane.
Thus, we shall begin with a Euclidean propagator, or a two-point Schwinger function $S (x-y) = \mathcal{S}_2(x,y) = \braket{\phi(x)\phi(y)}_{Euc.}$, assuming the ``temperedness'' condition and Euclidean invariance.
With some regularity assumption of $S(\xi)$ at $\xi = 0$, namely $S(\xi) \in \mathscr{S}'(\mathbb{R}^4)$, the Schwinger function can be expressed as
\begin{align}
    S(\xi) = \int \frac{d^4 k}{(2 \pi)^4}~ e^{ik \xi} D (k^2).
\end{align}
In the usual case where the K\"all\'en-Lehmann spectral representation holds, $D (k^2)$ has singularities only on the negative real axis, which is called the \textit{timelike axis}.
We call singularities except on the timelike axis \textit{complex singularities}.
A typical example of these singularities is a pair of complex conjugate poles as illustrated in Fig.~\ref{fig:complex.pdf}.

For technical reasons, we assume: (1) boundedness of complex singularities in $|k^2|$ in the complex $k^2$ plane, (2) holomorphy of $D(k^2)$ in a neighborhood of the real axis except for the timelike singularities, (3) some regularity of discontinuity on the timelike axis \footnote{Without this condition, the discontinuity would be, in general, a hyperfunction, which is inconvenient for some necessary limiting operations. Therefore, we assume that the discontinuity can be represented as a tempered distribution on $[-\infty,0]$, namely, $D(-\sigma^2 - i \epsilon) - D(-\sigma^2 + i \epsilon) \xrightarrow{\epsilon \rightarrow +0} \operatorname{Disc} D(-\sigma^2) \in \mathscr{S}'([0,\infty]), $
where $\mathscr{S}'([0,\infty])$ is the dual space of $\mathscr{S}([0,\infty]) := $ $\left\{ f(\lambda) = g(- (1+\lambda)^{-1}) ~;~g\mathrm{~is~a~}C^\infty \mathrm{~function~on~}[-1,0] \right\}$. For details, see Sec.~A.3 in \cite{Bogolyubov:1990kw}. Note that a tempered distribution with positive support can be extended to a tempered distribution on $[0,\infty]$.}, and (4) $D(k^2) \rightarrow 0$ as $|k^2| \rightarrow \infty$.

The following list outlines the main results of our study.
\begin{enumerate}
\renewcommand{\labelenumi}{(\Alph{enumi})}
    \item The reflection positivity
    is violated for the Schwinger function.
    \item The holomorphy of the Wightman function $W(\xi - i \eta)$ in the tube $\mathbb{R}^4 - i V_+$ and the existence of the boundary value as a distribution $W(\xi) := \lim_{\substack{\eta \rightarrow 0 \\ \eta \in V_+}} W (\xi - i \eta) \in \mathscr{D}'(\mathbb{R}^4)$ are still valid. Thus, we can reconstruct the Wightman function from the Schwinger function.
    \item The temperedness and the positivity condition in $\mathscr{D}(\mathbb{R}^4)$ are violated for the reconstructed Wightman function. The spectral condition is never satisfied since it requires the temperedness as a prerequisite.
    \item The Lorentz symmetry and spacelike commutativity are kept intact.
\end{enumerate}
We prove the assertions (A) -- (D) rigorously in \cite{HK2021} and provide essential ideas in this paper.
From the assertion (C), complex singularities may seem to have no physical interpretation. However, we argue:
\begin{enumerate}
\renewcommand{\labelenumi}{(\Alph{enumi})}
\setcounter{enumi}{4}
    \item Complex singularities can be realized in indefinite-metric QFTs and correspond to pairs of zero-norm eigenstates of complex energies.
\end{enumerate}

To demonstrate these results, we first give sketches of proofs of the assertions (A) -- (D) for an important example in the next section. We then mention their generalization to arbitrary complex singularities and discuss a possible realization in quantum theory.

\section{Special case: one pair of complex conjugate poles}

We begin with a propagator $D(k^2)$ with one pair of complex conjugate simple poles, which is decomposed into the ``timelike part'' $D_{tl}(k^2) $ and ``complex-pole part'' $D_{cp}(k^2)$,
\begin{align}
    D(k^2) &= D_{tl}(k^2)  + D_{cp}(k^2) , \notag \\
    D_{tl}(k^2) &= \int_0 ^\infty d \sigma^2 \frac{\rho(\sigma^2)}{\sigma^2 + k^2} , \notag \\
    D_{cp}(k^2) &= \frac{Z}{M^2 + k^2}  + \frac{Z^*}{(M^*)^2 + k^2}, \label{eq:propagator_complex_poles}
\end{align}
where $\rho(\sigma^2)$ is the spectral function and $M^2 \in \mathbb{C}$ is the complex mass squared.
Without loss of generality, we can choose $\operatorname{Im} M^2 > 0$.
Such a pair of complex conjugate poles is a typical example of complex singularities, which is suggested to appear in the transverse part of the Landau-gauge gluon propagator in many works \cite{Gribov78, Zwanziger89, Stingl85, HKRSW90, Stingl96, Dudal:2008sp, BT2019, Falcao:2020vyr,Siringo16a, Siringo16b, HK2018, HK2020,Fischer-Huber}.
Note that complex conjugate pairing of $D_{cp}(k^2)$ is necessary for a real field due to the Schwarz reflection principle.

We accordingly decompose the Schwinger function as
\begin{align}
    S(\xi) &= S_{tl}(\xi)  + S_{cp}(\xi), \notag \\
    S_{tl}(\xi) &= \int \frac{d^4 k}{(2 \pi)^4} e^{ik \xi} D_{tl}(k^2), \notag \\
    S_{cp}(\xi) &= \int \frac{d^4 k}{(2 \pi)^4} e^{ik \xi} D_{cp}(k^2).
\end{align}

We proceed to reconstruct the corresponding Wightman function as an analytic continuation of the Schwinger function by identifying Wightman function at pure imaginary time with the Schwinger function,
\begin{align}
    W(- i \xi_4, \vec{\xi}) := S(\vec{\xi} , \xi_4), 
\end{align}
for $\xi_4 > 0$.

\textbf{(B)} As usual, for the timelike part, we can analytically continue $W_{tl} (- i \eta^0, \vec{\xi}) = S_{tl} (\vec{\xi} , \eta^0)$ to the tube $\xi - i \eta = (\xi^0 - i \eta^0, \vec{\xi} - i \vec{\eta}) \in \mathbb{R}^4 - iV_+$.
Moreover, the limit $\eta \rightarrow 0$ in $\eta \in V_+$, or ``the boundary value,'' can be taken as a tempered distribution [in $\mathscr{S}'(\mathbb{R}^4)$],
\begin{align}
    W_{tl}(\xi^0,\vec{\xi}) &:= \lim_{\substack{\eta \rightarrow 0 \\ \eta \in V_+}} W_{tl} (\xi - i \eta) \notag \\
    &= \int_0^\infty d\sigma^2 ~ \rho(\sigma^2) i \Delta^+ (\xi, \sigma^2), \label{eq:timelike-Wightman}
\end{align}
which is formally a sum of the free Wightman function $i \Delta^+ (\xi, \sigma^2)$ of mass $\sigma^2$ with the weight $\rho (\sigma^2)$, where
\begin{align}
    i \Delta^+ (\xi, \sigma^2) &= (2 \pi) \int \frac{d^4 k}{(2 \pi)^4} e^{-i k \xi} \theta (k_0) \delta (k^2 - \sigma^2) 
\end{align}
with the Loretzian vectors $\xi = (\xi^0,\vec{\xi}), ~k = (k^0,\vec{k})$.

On the other hand, the complex-pole part $S_{cp}(\xi)$ can be expressed as
\begin{align}
    S_{cp}(\vec{\xi}, \xi_4) = \int \frac{d^3 \vec{k}}{(2 \pi)^3} e^{i\vec{k} \cdot \vec{\xi}} \left[ \frac{Z}{2 E_{\vec{k}} } e^{- E_{\vec{k}} |\xi_4|} +  \frac{Z^*}{2 E_{\vec{k}}^*} e^{- E_{\vec{k}}^* |\xi_4|}
    \right], \label{eq:simple_complex_poles_Schwinger}
\end{align}
where $E_{\vec{k}} = \sqrt{\vec{k}^2 +M^2}$ is a branch of $\operatorname{Re} E_{\vec{k}} > 0$ and $\operatorname{Im} E_{\vec{k}} > 0$ holds from the choice $\operatorname{Im} M^2 >0$.
Similarly to the timelike part, the complex-pole part of the Wightman function,
\begin{align}
    W_{cp}&(\xi - i\eta) = \int \frac{d^3 \vec{k}}{(2 \pi)^3} e^{i\vec{k} \cdot (\vec{\xi} - i \vec{\eta})} \notag \\
    &\times \left[ \frac{Z}{2 E_{\vec{k}} } e^{- i E_{\vec{k}} (\xi^0 - i \eta^0)} +  \frac{Z^*}{2 E_{\vec{k}}^*} e^{- i E_{\vec{k}}^* (\xi^0 - i \eta^0)} 
    \right]. \label{eq:simple_complex_poles_hol_Wightman}
\end{align}
is holomorphic in the tube $\mathbb{R}^4 - iV_+$, since the integrand decreases rapidly in $|\vec{k}|$ for $\xi - i \eta \in \mathbb{R}^4 - iV_+$. Note that $\operatorname{Im} M^2$ does not affect the convergence because of $E_{\vec{k}} = |\vec{k}| + O(1/|\vec{k}|)$.

We can regard the Fourier transform in (\ref{eq:simple_complex_poles_hol_Wightman}) as a tempered distribution in $\vec{\xi}$ with a smooth parameter $\xi^0$, which is a distribution in $\mathscr{D}'(\mathbb{R}^4)$. Then, the limit $\eta \rightarrow 0$ with $\eta \in V_+$ can be taken to yield the reconstructed Wightman function:
\begin{align}
    W_{cp}(\xi^0, \vec{\xi}) = \int \frac{d^3 \vec{k}}{(2 \pi)^3} e^{i\vec{k} \cdot \vec{\xi}} \left[ \frac{Z}{2 E_{\vec{k}} } e^{- i E_{\vec{k}} \xi^0} +  \frac{Z^*}{2 E_{\vec{k}}^*} e^{- i E_{\vec{k}}^* \xi^0}
    \right].
     \label{eq:simple_complex_poles_Wightman}
\end{align}
Therefore, we obtain the reconstructed Wightman function $W(\xi) = W_{tl}(\xi) + W_{cp}(\xi) $ as a distribution in $\mathscr{D}'(\mathbb{R}^4)$.

\textbf{(C)} 
Due to the exponential increases of the integrand as $\xi^0 \rightarrow \pm \infty$, $W_{cp}(\xi)$ is a nontempered distribution.
Thus, the reconstructed Wightman function is nontempered in the presence of complex poles $W(\xi) \notin \mathscr{S}'(\mathbb{R}^4)$.

Next, let us examine the positivity.
While the Wightman function is not a tempered distribution, we can still consider the positivity condition in $\mathscr{D}(\mathbb{R}^4)$, namely, for any test function with compact support $f \in \mathscr{D}(\mathbb{R}^4)$,
\begin{align}
    \int d^4 x d^4y~ W (y-x) f^*(x) f(y) \geq 0 .
    \label{eq:W-positivity}
\end{align}
This positivity condition ({\ref{eq:W-positivity}}) is violated due to the nontemperedness.

An intuitive derivation is as follows.
Intuitively, the positivity corresponds to that of the sector $\{ \phi(x) \ket{0} \}_{x \in \mathbb{R}^4}$. Suppose that this sector has a positive metric. From the translational invariance of the two-point function, the translation operator defined on the sector: $U(a) \phi(x) \ket{0} := \phi(x + a) \ket{0}$ is unitary. Since the modulus of a matrix element of a unitary operator is not more than one in a space with a positive metric, we have an $a$-independent ``upperbound,'' i.e., $|W(a)| = |\braket{0|\phi(0) U(-a) \phi(0)|0} | \leq \braket{0|\phi(0) \phi(0)|0}$.
This upperbound will imply that $W(a)$ is tempered, which contradicts the nontemperedness.
Of course, $W(0) = \braket{0|\phi(0) \phi(0)|0}$ is in general ill-defined since $W(\xi)$ is a distribution. A more delicate analysis is therefore required. We outline a rigorous proof in Appendix {\ref{app:appendix-pos}}.

\textbf{(A)} 
The reflection positivity is always violated with complex poles, since steps (a) and (b) of the OS theorem \cite{OS73} imply that the reflection positivity essentially yields the temperedness of the Wightman function.
We give a sketch of these steps in Appendix {\ref{app:appendix-pos}}.

\textbf{(D)} 
We consider the Lorentz symmetry and spacelike commutativity.
The invariance of the complex-pole part (\ref{eq:simple_complex_poles_hol_Wightman}) under Lorentz boosts can be explicitly checked by an integration path deformation.
Since the spatial rotational invariance is manifest, the Wightman function is Lorentz invariant. As another derivation, one can utilize the Euclidean invariance of the Schwinger function and the holomorphy in the tube. An argument similar to Bargmann-Hall-Wightman theorem (Theorem 2-11 and its Lemma of \cite{Streater:1989vi}) yields complex Lorentz invariance of $W(\xi - i \eta)$.
The spacelike commutativity is an immediate consequence of Lorentz invariance: $W(\xi) = W(-\xi)$ for spacelike $\xi$.

\section{General cases}

Let us mention a generalization of the above assertions to arbitrary complex singularities. With general complex singularities, the spectral representation is modified as, according to the Cauchy integral theorem,
\begin{align}
 D(k^2) &= \int_0 ^\infty d \sigma^2 \frac{\rho(\sigma^2)}{\sigma^2 + k^2} + \sum_{M} \oint_{\Gamma_M} \frac{d \zeta}{2 \pi i}  \frac{D(\zeta)}{\zeta - k^2}, \label{eq:spec_repr_complex}
 \end{align}
where $M$ is a label of a complex singularity and $\Gamma_M$ is a contour surrounding the singularity clockwise.
The contribution of each complex singularity is formally a sum of complex poles with with weight $- D(\zeta)/ 2 \pi i$ over the contour \footnote{The assumptions (1) boundedness of complex singularities in $|k^2|$ and (2) holomorphy of $D(k^2)$ in a neighborhood of the real axis except for the timelike singularities are required to justify this statement.}.
This leads to a generalization of the proof of (B) and (D).
We prove the nontemperedness of (C) as follows:
Suppose the Wightman function were tempered, then the holomorphy in the tube would essentially imply the spectral condition for the Wightman function in momentum representation.
This leads to the K\"all\'en-Lehmann spectral representation, which contradicts complex singularities.
Other claims of (A) and (C) follow from the nontemperedness as above.
For details, see \cite{HK2021}.

\section{Realization in quantum theory}

\textbf{(E)} 
We discuss a possible state-space structure.
Since abandoning the positivity of the full state space is common in Lorentz covariant gauge-fixed descriptions of gauge theories, we consider a quantum theory in a state space with an indefinite metric.
For a review on indefinite-metric QFTs, see e.g., \cite{Nakanishi72}.

Our aim here is to argue the correspondence between complex singularities and relevant complex-energy spectra. To this end, we shall demonstrate (E1) necessity and (E2) sufficiency of complex spectra for complex singularities when a convenient completeness relation is applicable.

\textbf{(E1)} 
We begin with the necessity of complex spectra for existence of complex singularities. 
Let us consider a $(0+1)$-dimensional QFT satisfying:
\begin{enumerate}
    \item completeness of denumerable eigenstates $\ket{n}$ of the Hamiltonian $H$: $1 = \sum_{n,n'} \eta^{-1}_{n,n'}  \ket{n} \bra{n'}$, where $\eta_{n,n'} = \braket{n|n'}$ is the nondegenerate metric \footnote{
Even in a finite dimensional vector space, the completeness of eigenstates of a hermitian operator does not always hold. Instead, by the Jordan decomposition, ``generalized eigenstates'' defined by a sequence $\{ \ket{E^{0}},~\ket{E^{1}},~\cdots, \ket{E^{n-1}} \}$: $ (H - E) \ket{E^{0}} = E \ket{E^{1}},~ (H - E) \ket{E^{1}} = E \ket{E^{2}},~\cdots~,~ (H - E) \ket{E^{n-1}} = 0$ spans the vector space.
Nevertheless, allowing generalized eigenstates does not change the conclusion.},
    \item translational covariance: $\phi(t) = e^{iHt} \phi(0) e^{-iHt}$,
    \item reality of eigenvalues $E_n$ of the Hamiltonian $H$.
\end{enumerate}
Then, with an assumption that the completeness relation converges well, the Wightman function is tempered as follows:
\begin{align}
    & \braket{0|\phi(t) \phi(0)|0} = \int d \omega ~ \rho (\omega) e^{-i\omega t}, \notag \\
    & \rho (\omega) = \sum_{n,n'} \eta^{-1}_{n,n'} \delta (\omega - E_n) \bra{0} \phi(0)  \ket{n} \bra{n'} \phi(0) \ket{0}.
\end{align}
This observation demonstrates that complex singularities do not appear without complex spectra, i.e., (E1) complex singularities require complex spectra of $H$.

\textbf{(E2)} 
On the other hand, eigenstates of complex eigenvalues of a Hermitian operator $H$ appear as pairs of zero-norm states. For example, a pair of zero-norm states $\{ \ket{\alpha}, \ket{\beta} \}$ can satisfy
\begin{align}
    \begin{cases}
    H \ket{\alpha} = E_\alpha \ket{\alpha}, ~~~ H \ket{\beta} = E^*_\alpha \ket{\beta} \\
    \braket{\alpha|\alpha} = \braket{\beta|\beta} = 0,~~ \braket{\alpha|\beta} \neq 0.
    \end{cases}
\end{align}

This pair of states $\{ \ket{\alpha}, \ket{\beta} \}$ can yield a pair of complex conjugate poles. Indeed, we find
\begin{align}
W_{complex}(t) &:= \sum_{n, n' \in \{ \alpha, \beta \} } \eta^{-1}_{n,n'} e^{- i E_n t} \bra{0} \phi(0)  \ket{n} \bra{n'} \phi(0) \ket{0} \notag \\
&= (\braket{\beta|\alpha})^{-1}  \braket{0|\phi(0)|\alpha} \braket{\beta|\phi(0)|0} e^{- i E_\alpha t} \notag \\
& + (\braket{\alpha|\beta})^{-1} \braket{0|\phi(0)|\beta} \braket{\alpha|\phi(0)|0}  e^{- i E_\alpha^* t},
\end{align}
which yields,
\begin{align}
    S_{complex}(\tau) &= W_{complex} (-i |\tau|) \notag \\ 
    &= \int \frac{dk}{2 \pi} e^{ik \tau} \left[ \frac{Z}{k^2 + E_\alpha^2} +  \frac{Z^*}{k^2 + (E_\alpha^*)^2} \right],
\end{align}
 with $Z := \frac{2 E_\alpha \braket{0|\phi(0)|\alpha} \braket{\beta|\phi(0)|0}}{\braket{\beta|\alpha}}$ for $\operatorname{Re} E_\alpha > 0$.
This indicates that (E2) a pair of zero-norm states $\{ \ket{\alpha}, \ket{\beta} \}$ with complex conjugate energies $\{ E_\alpha, E_\alpha^* \}$ satisfying $\operatorname{Re} E_\alpha > 0$ yields a pair of complex conjugate poles if $\{ \ket{\alpha}, \ket{\beta} \}$ are not orthogonal to $\phi(0) \ket{0}$.

These claims [(E1) and (E2)] establish the correspondence between complex singularities and pairs of zero-norm states of complex eigenvalues of the Hamiltonian.

\section{Example}

The above argument is in $(0+1)$-dimensional QFT.
We show an example of $(3+1)$-dimensional QFT yielding complex poles based on a covariant operator formulation \cite{Nakanishi72b} of the Lee-Wick model \cite{LW69}.
The Lagrangian density of a complex scalar field $\phi$ with complex (squared) mass $M^2 \in \mathbb{C}$ is given by
\begin{align}
\mathscr{L} :=& \frac{1}{2} \bigl[ (\partial_\mu \phi)(\partial^\mu \phi) + (\partial_\mu \phi)^\dagger (\partial^\mu \phi)^\dagger \notag \\
&~~~ - M^2 \phi^2 - (M^*)^2 (\phi^\dagger)^2  \bigr].
\end{align}
We expand the field operator $\phi$ as 
\begin{align}
\phi(x) &= \phi^{(+)} (x) + \phi^{(-)} (x), \notag \\
\phi^{(+)} (x) &= \int \frac{d^3 \vec{p}}{(2 \pi)^3} \frac{1}{\sqrt{2 E_{\vec{p}}}} \alpha(\vec{p}) e^{i \vec{p}\cdot \vec{x} - i E_{\vec{p}}t}, \notag \\
\phi^{(-)} (x) &= \int \frac{d^3 \vec{p}}{(2 \pi)^3} \frac{1}{\sqrt{2 E_{\vec{p}}}} \beta^\dagger(\vec{p}) e^{-i \vec{p}\cdot \vec{x} + i E_{\vec{p}}t},
\end{align}
where $E_{\vec{p}} := \sqrt{M^2 + \vec{p}^2}$, $\operatorname{Re} E_{\vec{p}} \geq 0$, and $\operatorname{Re} \sqrt{E_{\vec{p}}} \geq 0$.
The canonical commutation relation implies $[\alpha(\vec{p}), \beta^\dagger(\vec{q})] =[\beta(\vec{p}), \alpha^\dagger(\vec{q})] = (2 \pi)^3 \delta(\vec{p} - \vec{q})$.
We define the vacuum $\ket{0}$ by $\phi^{(+)} (x) \ket{0} = [\phi^{(-)} (x) ]^\dagger \ket{0} = 0$, which is a Lorentz invariant state, see \cite{Nakanishi72b} for details.
Note that one can explicitly check the spacelike commutativity at least at the level of elementary fields.

The Hamiltonian reads,
\begin{align}
H = \int \frac{d^3 \vec{p}}{(2 \pi)^3} \left[ E_{\vec{p}} \beta^\dagger(\vec{p}) \alpha(\vec{p}) + E_{\vec{p}}^* \alpha^\dagger(\vec{p}) \beta(\vec{p})  \right],
\end{align}
up to some constant. The complex-energy states $\ket{\vec{p},\alpha} := \alpha^\dagger(\vec{p}) \ket{0}$ and $\ket{\vec{p},\beta} := \beta^\dagger(\vec{p}) \ket{0}$ form a pair of zero-norm states:
$\braket{\vec{p},\alpha|\vec{q},\alpha} = \braket{\vec{p},\beta|\vec{q},\beta} = 0, ~~ \braket{\vec{p},\alpha|\vec{q},\beta} = \braket{\vec{p},\beta|\vec{q},\alpha} = (2 \pi)^3 \delta(\vec{p} - \vec{q}).$

We find that the Euclidean propagator of a Hermitian combination with a constant $Z \in \mathbb{C}$, $\Phi := \sqrt{Z} \phi + \sqrt{Z^*} \phi^\dagger$,
has complex poles. Indeed, the Wightman function of the Lee-Wick model,
\begin{align}
W_\Phi (t,\vec{x}) &:= \braket{0|\Phi(x)\Phi(0)|0} \notag \\
&= \int \frac{d^3 \vec{p}}{(2 \pi)^3} \left[ \frac{Z}{ 2 E_{\vec{p}}}  e^{i \vec{p}\cdot \vec{x} - i E_{\vec{p}}x^0} + \frac{Z^*}{ 2 E_{\vec{p}}^*}  e^{i \vec{p}\cdot \vec{x} - i E_{\vec{p}}^* t} \right],
\end{align}
coincides with the Wightman function (\ref{eq:simple_complex_poles_Wightman}) reconstructed from a pair of simple complex conjugate poles.

\section{Concluding Remarks}

Some remarks on the nontemperedness, locality, and Wick rotation are in order.

\subsection{Nontemperedness}
The exponential growth of the Wightman function $W(\xi)$ largely affects asymptotic states, which correspond to ``$\xi^0 \rightarrow \pm \infty$ limit''. This indicates that asymptotic states of the field are ill-defined without some artificial manipulations.
Such states in the full state space are far from being identified with asymptotic particle states and should be excluded from the physical state space before taking the asymptotic limit through, e.g. the Kugo-Ojima quartet mechanism \cite{Kugo:1979gm}.
Thus, \textit{the complex singularities can be considered as a signal of confinement}.
Incidentally, note that the appearance of complex singularities in a propagator of the gluon-ghost composite operator is a necessary condition for eliminating complex-energy states in ``the one-gluon state'' from the physical state space in the Becchi-Rouet-Stora-Tyutin (BRST) formalism.
Seeking such complex gluon-ghost bound states would be interesting for future prospects.
Remarkably, the Bethe-Salpeter equation for the gluon-ghost bound state has been discussed in light of BRST quartets in {\cite{Alkofer:2011pe}}.


\subsection{Locality}
Some argue that complex singularities are associated with nonlocality.  For example, it is claimed in \cite{Stingl85,Stingl96,HKRSW90} that complex poles describe short-lived excitations and that the locality is broken in short range at the level of propagators but that the corresponding $S$ matrix remains causal. However, this interpretation is different from ours.
To our knowledge, the only axiomatic way to impose locality is the spacelike commutativity.
Consequently, \textit{complex singularities themselves not necessarily lead to non-locality}, as shown in this paper.

\subsection{Wick rotation}
Our reconstruction procedure is different from the naive inverse Wick rotation in the momentum space $k_E^2 \rightarrow - k^2$. Indeed, due to the complex singularities, the time-ordered propagator cannot be Fourier transformed. On the other hand, the inverse Wick rotation makes the time-ordered propagator tempered and ruins the interpretation of Euclidean field theory as an imaginary-time formalism.
Note also that the inverse Wick rotation invalidate the Hermiticity of the Hamiltonian even in an indefinite space unlike ours.

In summary, complex singularities are beyond the standard formulation yet consistent with locality. Moreover, they can appear in indefinite-metric QFTs, e.g., gauge theories in Lorentz covariant gauges.

\section*{Acknowledgements}
We thank Taichiro Kugo, Peter Lowdon, and Lorenz von Smekal for helpful and critical comments in the early stage of this work.
Y.~H. is supported by JSPS Research Fellowship for Young Scientists Grant No.~20J20215, and K.-I.~K. is supported by Grant-in-Aid for Scientific Research, JSPS KAKENHI Grant (C) No.~19K03840.

\appendix

\section{Outlines of proofs for positivity violations} \label{app:appendix-pos}
We outline proofs for violations of the positivity (\ref{eq:W-positivity}) and reflection positivity, by showing how temperedness arises from each of the positivity conditions.

\subsection{Violation of the positivity (\ref{eq:W-positivity})}
Suppose that the positivity ({\ref{eq:W-positivity}}) holds.
We define a positive semidefinite sesquilinear form on $\mathscr{D}(\mathbb{R}^4)$: for $f,g \in \mathscr{D}(\mathbb{R}^4)$,
\begin{align}
    (f,g)_W := \int d^4 x d^4y~ W (y-x) f^*(x) g(y). 
\end{align}
For $a \in \mathbb{R}^4$, $\hat{U}(a)$ denotes the translation operator on $\mathscr{D}(\mathbb{R}^4)$ satisfying $(\hat{U}(a) f, \hat{U}(a) f)_W = (f,f)_W$.
Since $(\cdot,\cdot)_W$ is positive semidefinite, the Cauchy-Schwarz inequality yields a bound on $( f * (g * W))(a)$: $|( f * (g * W))(a)| = |(f^*, \hat{U}(a) \hat{g} )_W| \leq \sqrt{ (f^*,f^*)_W  (g,g)_W  }$ for all $f,g \in \mathscr{D}(\mathbb{R}^4)$, where $\hat{g} (x) := g(-x)$ and $*$ denotes the convolution.
Note that $T \in \mathscr{D}'(\mathbb{R}^4)$ is tempered if and only if $\alpha * T$ is a smooth function of at most polynomial growth for any $\alpha \in \mathscr{D}(\mathbb{R}^4)$ (see Theorem 6, Chapter 7 in \cite{Schwartz}). Using this criterion to $( f * (g * W))(a)$ and $(g * W)(x)$, we obtain $W \in \mathscr{S}'(\mathbb{R}^4)$, which contradicts the nontemperedness.

\subsection{Violation of the reflection positivity}
Let us sketch out a proof of the reflection-positivity violation.
It is sufficient to show the violation of the reflection positivity 
for the two-point function, which is a necessary condition for the reflection positivity: for any test function with positive support $f \in \mathscr{S}(\mathbb{R}^4_+) := \{ f \in \mathscr{S} (\mathbb{R}^4) ~;~ \operatorname{supp} f \subset \mathbb{R}^3 \times [0,\infty) \}$,
    \begin{align}
    \int d^4x d^4y ~ f^* (\vartheta x) f(y) S (x-y) \geq 0, \label{eq:two_pt_ref_pos}
\end{align}
where $\vartheta$ is the reflection defined by $\vartheta x := (\vec{x},-x_4)$.

Suppose that this condition ({\ref{eq:two_pt_ref_pos}}) holds. Then, the positivity ({\ref{eq:two_pt_ref_pos}}) naturally provides a positive semidefinite sesquilinear form on $\mathscr{S}(\mathbb{R}^4_+)$. By dividing $\mathscr{S}(\mathbb{R}^4_+)$ by the zero-norm subspace, $\mathscr{N}$ and completing the pre-Hilbert space, we obtain a Hilbert space $\mathscr{K} = \overline{\mathscr{S}(\mathbb{R}^4_+)/\mathscr{N}}$. In this Hilbert space, the imaginary-time translation operators $\{ T^\tau \}$ form a semigroup of self-adjoint operators satisfying $\| T^\tau \varphi \| \leq \| \varphi \|$ for any $ \varphi \in \mathscr{K}$. Thus, we can represent $T^\tau = e^{-\tau H}$ with a self-adjoint generator $H$. Using the holomorphic semigroup $T^{\tau +is} = e^{-(\tau +is) H}$ on $(0,\infty) \times \mathbb{R}$, we can construct an analytic continuation of $S(\xi)$ (smeared in the spatial directions). In the real-time direction, this analytic continuation is tempered since $T^{is}$ is unitary and thus bounded in $s$. This leads to the temperedness of the Wightman function, which contradicts the nontemperedness.

\end{document}